\documentclass[aps,pra,twocolumn,superscriptaddress]{revtex4-1}

\usepackage{graphicx}
\usepackage[caption=false]{subfig}

\usepackage{amsmath, amsfonts, amssymb}
\usepackage{nopageno, tikz, mathrsfs, xcolor, hyperref, upgreek, dcolumn, bm}

\hypersetup{colorlinks, linkcolor=teal, citecolor=blue, urlcolor=purple}

\usetikzlibrary{decorations.markings}

\makeatletter
\def\DC@endright{$\hfil\egroup\@dcolcolor\box\z@\box\tw@\dcolreset}
\def\dcolcolor#1{\gdef\@dcolcolor{\color{#1}}}
\def\dcolreset{\dcolcolor{black}}
\dcolcolor{black}
\makeatother

\newcommand{\currentLR}{\hspace{-0.6ex}\begin{tikzpicture}[baseline=0.15ex]
\draw[-,thin,decoration={markings, mark=at position 1 with {\arrow{>}}},decorate] (-1.5ex,0.75ex) -- (-0.2ex,1.5ex);
\draw[-,thin,decoration={markings, mark=at position 1 with {\arrow{>}}},decorate] (1.5ex,0.75ex) -- (0.2ex,1.5ex);
\draw[-,thin] (0,0) arc[radius = 0.75ex, start angle = -90, end angle = -250];
\draw[-,thin] (0,0) arc[radius = 0.75ex, start angle = -90, end angle = 70];
\end{tikzpicture}\hspace{-0.5ex}}

\newcommand{\currentR}{\hspace{-0.6ex}\begin{tikzpicture}[baseline=0.15ex]
\draw[-,thin,decoration={markings, mark=at position 1 with {\arrow{>}}},decorate] (-1.5ex,0.75ex) -- (-0.2ex,1.5ex);
\draw[-,thin] (0,0) arc[radius = 0.75ex, start angle = -90, end angle = -250];
\draw[-,thin] (0,0) arc[radius = 0.75ex, start angle = -90, end angle = 70];
\end{tikzpicture}\hspace{0.2ex}}

\newcommand{\currentL}{\hspace{0.1ex}\begin{tikzpicture}[baseline=0.15ex]
\draw[-,thin,decoration={markings, mark=at position 1 with {\arrow{>}}},decorate] (1.5ex,0.75ex) -- (0.2ex,1.5ex);
\draw[-,thin] (0,0) arc[radius = 0.75ex, start angle = -90, end angle = -250];
\draw[-,thin] (0,0) arc[radius = 0.75ex, start angle = -90, end angle = 70];
\end{tikzpicture}\hspace{-0.5ex}}

\newcommand{\blank}{\,\text{\raisebox{0.4ex}{$\scriptscriptstyle \bullet$}}\,}
\newcommand{\Blank}{\,\text{\raisebox{0.4ex}{$\scriptscriptstyle \circ$}}\,}

\newcommand{\T}{\rule{0pt}{2.6ex}}       
\newcommand{\B}{\rule[-1.2ex]{0pt}{0pt}} 

\begin{document}

\title{Superconducting circuit protected by two-Cooper-pair tunneling}

\author{W.\ C.\ Smith}
\thanks{Current Address: QUANTIC Team, INRIA Paris, 2 rue Simone Iff, 75012 Paris, France}
\email[]{william.smith@inria.fr}
\author{A.\ Kou}
\author{X.\ Xiao}
\author{U.\ Vool}
\author{M.\ H.\ Devoret}
\email[]{michel.devoret@yale.edu}
\affiliation{Departments of Applied Physics and Physics, Yale University, New Haven, CT 06520, USA}

\date{\today}

\begin{abstract}

We present a protected superconducting qubit based on an effective circuit element that only allows pairs of Cooper pairs to tunnel. These dynamics give rise to a nearly degenerate ground state manifold indexed by the parity of tunneled Cooper pairs. We show that, when the circuit element is shunted by a large capacitance, this manifold can be used as a logical qubit that we expect to be insensitive to multiple relaxation and dephasing mechanisms.

\end{abstract}


\maketitle

\section{Introduction\label{sec:intro}}

\subsection{Motivation\label{subsec:motivation}}

Superconducting circuits are widely recognized as a powerful potential platform for quantum computation and now stand at the frontier of quantum error correction \cite{Devoret2013}. Future progress will likely stem from two complementary strategies: (i) active error correction characterized by measurement-based \cite{Kelly2015,Riste2015,Ofek2016} and autonomous \cite{Murch2012,Shankar2013,Leghtas2015,Puri2017} stabilization, and (ii) passive error correction characterized by protected qubits \cite[and references therein]{Doucot2012}. We address strategy (ii) in this article by designing an experimentally accessible protected qubit.

\begin{figure}
\centering
\subfloat{%
\includegraphics{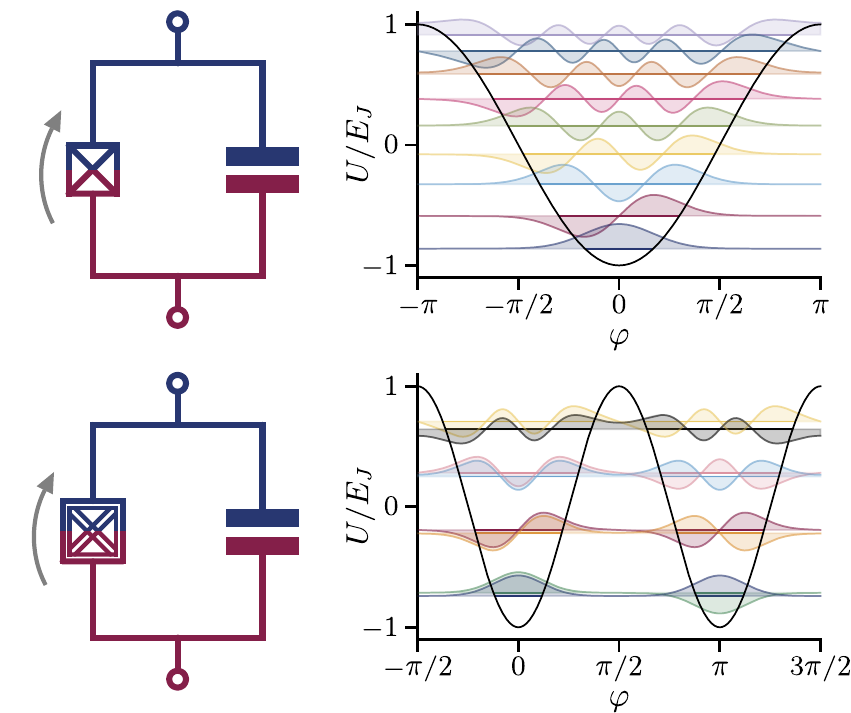}
\put(-245,198){(a)}
\put(-149,198){(b)}
\put(-245,93){(c)}
\put(-149,93){(d)}
\put(-244,50){$\varphi$}
\put(-242,155){$\varphi$}
\put(-215,64){$E_J$}
\put(-215,167){$E_J$}
\put(-196,145){$C_\text{shunt}$}
\put(-196,40){$C_\text{shunt}$}
\label{fig:transmon}}
\subfloat{
\label{fig:t_potential}}
\subfloat{
\label{fig:cos2phi}}
\subfloat{
\label{fig:c_potential}}
\caption{(a) Electrical circuit for the transmon qubit. (b) Potential energy of the transmon with energy levels and wavefunctions for the first few eigenstates. (c) Electrical circuit for the idealized protected qubit. The cross-hatched circuit element comprises a capacitance in parallel with an inductive element that exclusively permits the tunneling of pairs of Cooper pairs. The superconducting island is indicated by color. (d) Potential energy of the ideal charge-protected qubit with the lowest energy levels and wavefunctions.}
\end{figure}
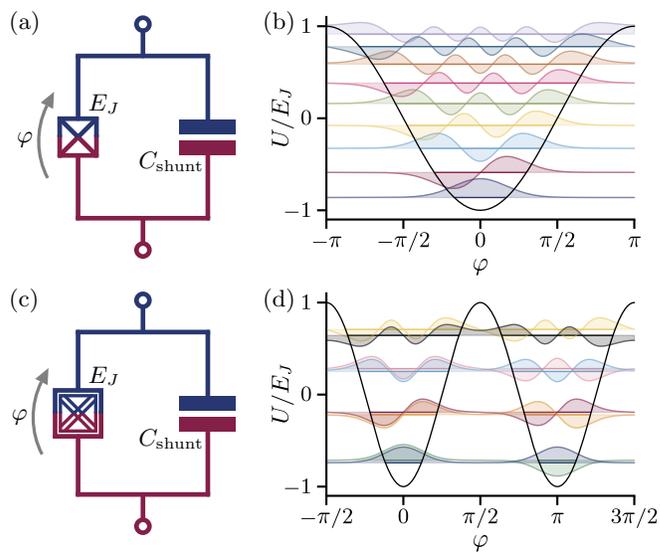

The transmon qubit \cite{Koch2007.1} has proven to be a remarkably successful prototype of a protected qubit. Its circuit (depicted in Fig.\ \ref{fig:transmon}) contains a Josephson junction whose potential energy is $U = -E_J \cos \varphi$, with $E_J$ being the tunneling energy and $\varphi$ being the superconducting phase across the junction. The large shunt capacitance imposes a large effective mass for the analogous ``particle in a box,'' confining the low-energy wavefunctions near $\varphi = 0$, the only minimum for $\varphi \in (-\pi, \pi)$. This confinement suppresses the susceptibility of the qubit to offset charge noise and renders the energy spectrum approximately harmonic with level spacing much smaller than $E_J$ (see Fig.\ \ref{fig:t_potential}). On the other hand, circuit elements with degenerate phase states that only allow tunneling of pairs of Cooper pairs, meaning their potential energy is $U = - E_J \cos 2\varphi$, have been developed in recent years as a building block for topologically protected qubits \cite{Doucot2002,Gladchenko2009}. In this article, we propose a transmon-like qubit with additional protection from environmental noise by combining the large shunt capacitance of the transmon with such a $\cos2\varphi$ circuit element (the cross-hatched box in Fig.\ \ref{fig:cos2phi}). In this case, the wavefunctions are localized near $\varphi = 0, \pi$ (see Fig.\ \ref{fig:c_potential}), resulting in a nearly degenerate harmonic level arrangement. While the detrimental effects of offset charge noise are similarly suppressed in this circuit, sensitivity of the qubit to other decoherence mechanisms is also reduced, owing to the conservation of Cooper pair number parity.

In this article, we introduce a few-body transmon-type qubit where the charge carriers are exclusively pairs of Cooper pairs. Our central result is that there exists an experimentally attainable parameter regime for which conservative predictions of relaxation and dephasing times exceed $1\,$ms, i.e.\ an order of magnitude higher than those of typical transmons, given the same environmental noise \cite{Rigetti2012,Wang2019}. In the remainder of Sec.\ \ref{sec:intro}, we describe a toy model for the protected qubit. We proceed by analytically and numerically examining the Hamiltonian for the full superconducting circuit in Sec.\ \ref{sec:circuit}. Our attention in Sec.\ \ref{sec:qubit} then turns to properties of the ground state manifold, which we envision using as a protected qubit. A brief discussion about the concept of protection and examples of protected qubits, as well as our perspectives on readout and control, follows in Sec.\ \ref{sec:discussion}. Finally, we summarize our results in Sec.\ \ref{sec:conclusion}.

\subsection{\boldmath$\cos 2\varphi$ element\label{subsec:element}}

We first examine the advantages of the ideal circuit in Fig.\ \ref{fig:cos2phi} as a protected qubit. This circuit can be viewed as a Josephson-junction-like element (the cross-hatched box) shunted by a capacitance. Pairs of Cooper pairs are the only charge excitations permitted to tunnel through this element \cite{Doucot2002}. In the Cooper pair number basis, the potential energy assumes the form
\begin{equation*}
-\frac{1}{2} E_J \sum\limits_{N=-\infty}^\infty \left( |N\rangle \langle N+2| + |N+2\rangle \langle N| \right) = -E_J \cos 2\varphi,
\end{equation*}
where $E_J$ is the effective tunneling energy of the process. This expression follows from the conjugacy relation $[\varphi, N] = i$, where $N$ is the number of Cooper pairs that have tunneled. The invariance of the potential under translations in $\varphi$ by multiples of $\pi$ implies that half-fluxons are able to traverse the element.

The shunt capacitance and other charging effects produce a quadratic kinetic energy, yielding the Hamiltonian
\begin{equation}
H = 4 E_C (N - N_\text{g})^2 - E_J \cos 2\varphi, \label{eq:simple_ham}
\end{equation}
where $E_C$ is the charging energy and $N_\text{g}$ is the offset charge. This offset charge has been introduced due to the periodicity of the Hamiltonian in $\varphi$, which reflects the presence of a superconducting island in the circuit (as colored in Fig.\ \ref{fig:cos2phi}).

Since the circuit element only allows pairs of Cooper pairs to tunnel, the parity of the number of Cooper pairs that have tunneled is preserved under the action of the Hamiltonian. This property leads to two nearly degenerate ground states $|+\rangle$ and $|-\rangle$, which only consist of even and odd Cooper pair number states, respectively \cite{Gladchenko2009}. Since these states have no overlap in charge space (equivalently, they have opposite periodicity in phase space---see Fig.\ \ref{fig:c_potential}), we have $\langle - | \mathcal{O} | + \rangle \approx 0$ for any sufficiently local operator $\mathcal{O}$. Furthermore, the states \footnote{This notation is chosen for consistency with Sec.\ \ref{sec:circuit}, where there is a correspondence to persistent current handedness.} $|\currentLR\rangle = \frac{1}{\sqrt{2}}(|+\rangle \pm |-\rangle)$ are respectively localized near $\varphi = 0, \pi$ (see Fig.\ \ref{fig:c_potential}). Because these states have suppressed overlap in phase space for large $E_J/E_C$ (i.e.\ they are roughly inversely periodic in charge space), we have $\langle \currentL | \mathcal{O} | \currentR \rangle \approx 0$ for similarly local $\mathcal{O}$. These are precisely the conditions for simultaneously suppressing spurious transitions and phase changes between the states \cite{Ioffe2002.1} [resembling a Gottesman-Kitaev-Preskill (GKP) encoding on a circle \cite{Gottesman2001}].

More concretely, the ground state splitting obeys
\begin{equation*}
\Delta E \approx 16 E_C \sqrt{\frac{2}{\pi}} \left(\frac{2E_J}{E_C}\right)^{3/4} \text{e}^{-\sqrt{2E_J/E_C}} \cos (\pi N_\text{g}),
\end{equation*}
for large $E_J/E_C$ (see App.\ \ref{app:mathieu}). The two ground state energies oscillate out of phase with one another in $N_\text{g}$. Moreover, this shows that the splitting, as well as the charge dispersion, is exponentially suppressed in $E_J/E_C$. Thus, the role of the shunt capacitance is to decrease the charging energy $E_C$ and hence mitigate offset charge noise, much like in the transmon qubit \cite{Koch2007.1}.

\section{Superconducting circuit\label{sec:circuit}}

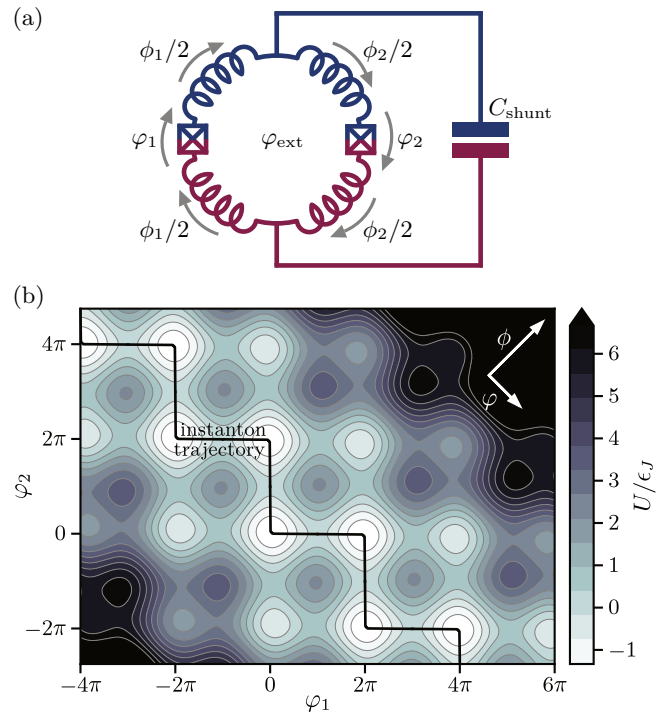
\begin{figure}[b]
\centering
\subfloat{%
\includegraphics{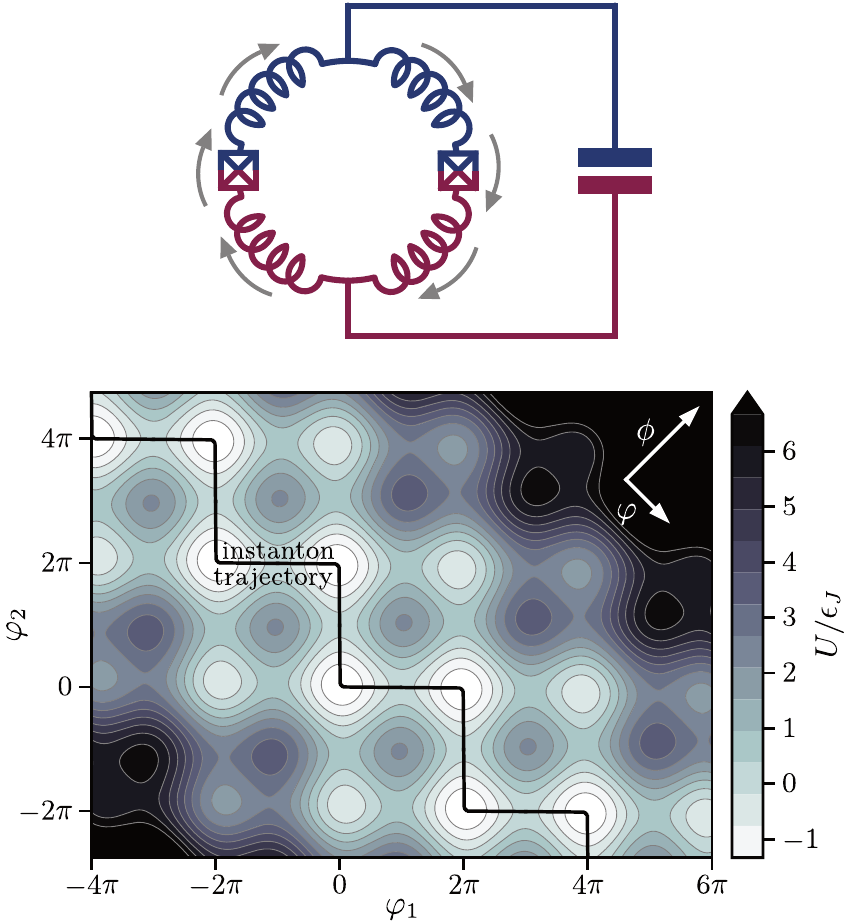}
\put(-245,260){(a)}
\put(-245,155){(b)}
\put(-151,215){$\varphi_\text{ext}$}
\put(-200,215){$\varphi_1$}
\put(-99.5,215){$\varphi_2$}
\put(-112.5,248){$\phi_2/2$}
\put(-112.5,180){$\phi_2/2$}
\put(-197,180){$\phi_1/2$}
\put(-197,248){$\phi_1/2$}
\put(-65,226){$C_\text{shunt}$}
\label{fig:circuit}}
\subfloat{\label{fig:instanton}}
\caption{(a) Reduced electrical circuit for the physical protected qubit. When $\varphi_\text{ext} = \pi$, the two Josephson junctions and superinductances collectively behave as the cross-hatched element. The superconducting island is indicated by color. (b) Contour plot of the potential energy $U$ in Eq.\ \ref{eq:sym_ham} in the $\varphi_1 \varphi_2$-plane at $\varphi_\text{ext} = \pi$ for $\theta = 0$. The numerically computed instanton trajectory between adjacent potential minima is overlaid in black. Importantly, this trajectory closely resembles a sequence of straight lines.}
\end{figure}

\subsection{Hamiltonian\label{subsec:hamiltonian}}

We now detail the superconducting circuit necessary to practically implement the sought-after $\cos 2 \varphi$ Josephson element. This circuit, depicted in Fig.\ \ref{fig:circuit}, is composed of two identical arms, each containing a Josephson junction in series with superinductances \cite{Manucharyan2009,Masluk2012}, arranged in parallel \cite{Kitaev2006,Brooks2013} and shunted by a large capacitance. These superinductances are split in half and placed on either side of the respective Josephson junctions to avoid large capacitances to ground. Kirchhoff's current law allows us to treat the phases across the superinductances in series as equal. When the external magnetic flux threading the inductive loop reaches half of a flux quantum, i.e.\ when $\varphi_\text{ext}=\pi$, the loop approximates the cross-hatched box in Fig.\ \ref{fig:cos2phi}. At this particular bias point, a Cooper pair can only tunnel through one of the Josephson junctions if it is accompanied by another Cooper pair tunneling through the other junction (in either direction). Conversely, a fluxon traversing a single Josephson junction corresponds to a half-fluxon traversing the whole element.

We consider the symmetric and antisymmetric combinations of superconducting phase coordinates
\begin{equation*}
\phi = \varphi_1 + \varphi_2 \qquad \varphi = \frac{1}{2}(\varphi_1 - \varphi_2) \qquad \theta = \frac{1}{2}(\phi_1 - \phi_2),
\end{equation*}
and their conjugate charges $\{n, N, \eta\}$ in numbers of Cooper pairs. Note that the prefactor in the definition of $\varphi$ is chosen to bring the coordinate into agreement with the phase drop across the inductive loop in the limit that $\theta$ vanishes. The Hamiltonian reads
%
%
\begin{align}
H &= 4 \epsilon_C \left[2 n^2 + \frac{1}{2} (N - N_\text{g} - \eta)^2 + x \eta^2\right] \nonumber \\
&\hspace{0.5cm}+ \epsilon_L \left[ \frac{1}{4}(\phi - \varphi_\text{ext})^2 + \theta^2 \right] - 2\epsilon_J\cos \varphi \cos \frac{\phi}{2}, \label{eq:sym_ham}
\end{align}
where $\epsilon_C$ and $\epsilon_J$ are the single junction charging and tunneling energies, $2\epsilon_L$ is the inductive energy of each superinductance, $x \equiv C_J / C_\text{shunt}$ is the ratio of the junction capacitance to the shunt capacitance, and $N_\text{g}$ is the offset charge on the superconducting island (see Fig.\ \ref{fig:circuit}). 

From this expression, it is clear that this circuit has three strongly coupled modes. The $\phi$ mode is flux dependent and is strongly and nonlinearly coupled, via the Josephson junctions, to the $\varphi$ mode. The $\varphi$ mode is offset charge dependent and strongly but linearly capacitively coupled to the $\theta$ mode. Our analysis and the effects observed in the remainder of this article require the parameter regime $\epsilon_L \ll \epsilon_J$, $\epsilon_C \lesssim \epsilon_J$, and $x \ll 1$ \footnote{Specifically, the theory in Sec.\ \ref{subsec:semiclassical} breaks down at $\epsilon_L \sim \epsilon_J$ and the protection in Sec.\ \ref{sec:qubit} breaks down for $x \gtrsim 0.1$.}. In particular, the parameters chosen for numerical simulations are listed in Tab.\ \ref{tab:params} and are similar to those of recent fluxonium devices \cite{Manucharyan2009,Earnest2018}.

\subsection{Semiclassical theory\label{subsec:semiclassical}}

In order to gain insight into the structure of the Hamiltonian in Eq.\ \ref{eq:sym_ham}, we briefly revisit its potential energy $U$ in the $\varphi_1\varphi_2$-plane, which is plotted in Fig.\ \ref{fig:instanton} for $\theta=0$. The cosine terms in the potential form a two-dimensional ``egg carton'' of wells. The minimum of the quadratic term in the potential occurs at $\varphi_1 + \varphi_2 = \varphi_\text{ext}$, which generally falls between adjacent diagonal ridges of cosine wells. At the special value of $\varphi_\text{ext} = \pi$, these two ridges are degenerate. Near this value of the external flux, we consider the path of the system between neighboring potential minima by numerically solving for the three-dimensional instanton \cite{Matveev2002} trajectory (see App.\ \ref{app:instanton}).

\begin{table}
\begin{ruledtabular}
\begin{tabular}{c c c c}
$\epsilon_J/h$ & $\epsilon_C/h$ & $\epsilon_L/h$ & $x$ \B \\
\hline
15 GHz & 2 GHz & 1 GHz & 0.02 \T \\
\end{tabular}
\end{ruledtabular}
\caption{Circuit parameters used for numerical simulations.\label{tab:params}}
\end{table}

We then constrain the system to this maximally probable tunneling path. For the parameters mentioned above, this path is well-described by
\begin{equation}
\phi = \frac{1}{1 + z} \left( 2 \left| \varphi - 2\pi \, \text{round}\, \frac{\varphi}{2\pi} \right| + z \varphi_\text{ext}\right),\label{eq:path}
\end{equation}
where $z \equiv \epsilon_L/\epsilon_J$ is a small parameter \footnote{The main discrepancy between the approximate path in Eq.\ \ref{eq:path} and the calculated path in Fig.\ \ref{fig:instanton} is the sharp bending near the minima, which arises due to coupling to the $\theta$ variable.}. Plugging this expression into the Hamiltonian in Eq.\ \ref{eq:sym_ham}, approximating the resulting one-dimensional potential by the first few terms in its Fourier series, and Taylor expanding about $z = 0$ yields $H \approx H_\text{eff}$ with
%
%
\begin{align}
H_\text{eff} &= 4 \epsilon_C \left[ \frac{1}{4 (1-z)} (N - N_\text{g} - \eta)^2 + x\eta^2\right] + \epsilon_L \theta^2 \nonumber\\
&\hspace{0.2cm} - \frac{16}{3\pi} \epsilon_L (\pi - \phi_\text{ext}) \cos \varphi - \epsilon_J \left(1 - \frac{5}{4} z\right)\cos 2\varphi \label{eq:2m_ham}
\end{align}
to leading order. Here, $\phi_\text{ext} = \left|\varphi_\text{ext} - 4\pi \, \text{round} \, \frac{\varphi_\text{ext}}{4\pi} \right|$ and we have discarded terms higher than the second harmonic or $O(\epsilon_L)$ (see App.\ \ref{app:instanton} for additional terms). This treatment exposes the ``$\cos 2\varphi$ nature'' of the potential at $\varphi_\text{ext} = \pi$, where the $\cos \varphi$ term vanishes. By comparison to Eq.\ \ref{eq:simple_ham}, we see that the added complication is that the $\varphi$ mode is strongly coupled to the $\theta$ mode. The resulting hybridization is a central ingredient to understanding properties of the system beyond the ground state manifold (see Sec.\ \ref{subsec:wavefunctions}).

We comment that this approximation neglects quantum fluctuations that are perpendicular to the path in Eq.\ \ref{eq:path}. This is consequently a semiclassical approximation: we have minimized the energy of the system with respect to the dynamical coordinate orthogonal to the trajectory \cite{Matveev2002}. Moreover, from Fig.\ \ref{fig:instanton}, it is clear that the approximation we have made is that fluxons traverse a single Josephson junction at a time.

\subsection{Energy spectrum\label{subsec:spectrum}}

\begin{figure*}
\centering
\subfloat{%
\includegraphics{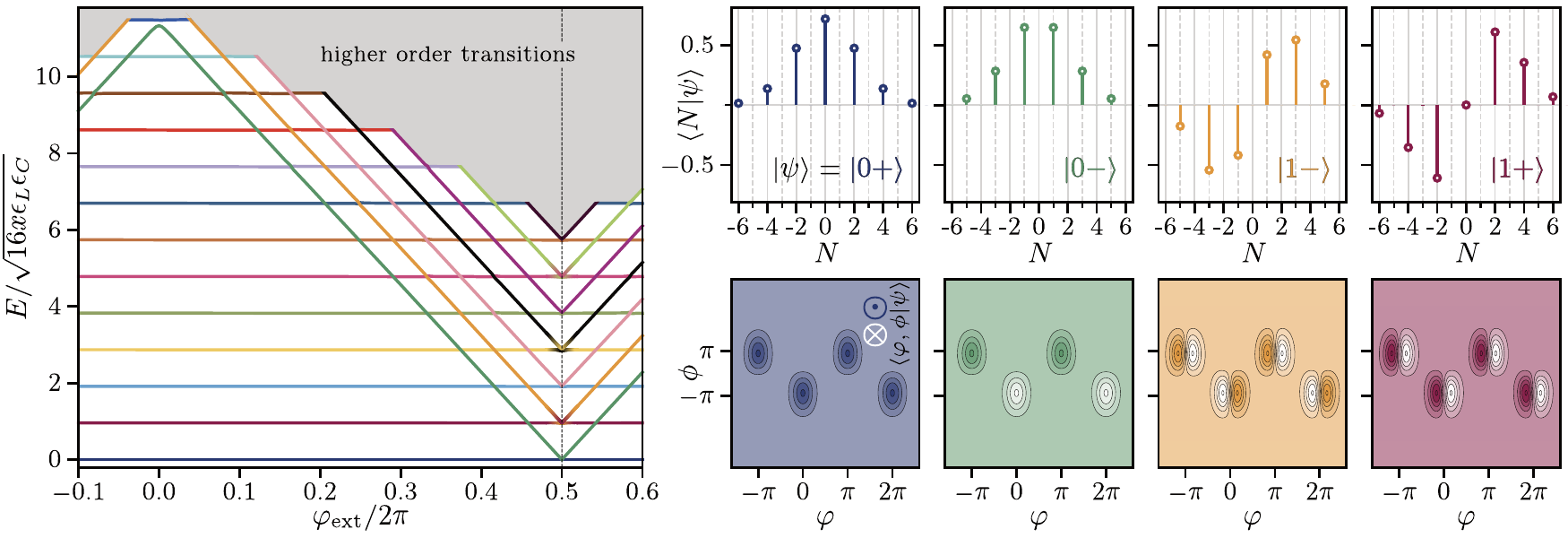}
\put(-507,170){(a)}
\put(-293,170){(b)}
\put(-293,83){(c)}
\label{fig:spectruma}}
\subfloat{\label{fig:spectrumb}}
\subfloat{\label{fig:spectrumc}}
\caption{(a) Normalized transition energies $E$ from the ground state of the Hamiltonian in Eq.\ \ref{eq:sym_ham} as a function of external flux at $N_\text{g} = 0$. The essential feature is the presence of a flux-independent plasmon mode (maroon) with energy $\sqrt{16 x \epsilon_L \epsilon_C}$ and a flux-dependent fluxon mode (green). The coloring reflects the assigned quantum numbers. (b) Charge wavefunctions $\langle N | \psi \rangle = \langle N | m \pm\rangle$ of the first four eigenstates of Eq.\ \ref{eq:sym_ham}, illustrating that the plasmon excitation index $m=0,1,\dots$ indicates the functional form while the fluxon excitation index $\pm$ indicates charge parity. (c) Phase wavefunctions $\langle \varphi, \phi | \psi \rangle$ of the first four eigenstates of Eq.\ \ref{eq:sym_ham}, showing that the intrawell excitation number corresponds to the plasmon index and that the bonding/antibonding configuration corresponds to the fluxon index. All wavefunctions are computed at $\varphi_\text{ext} = \pi$. \label{fig:spectrum}}
\end{figure*}

From numerical diagonalization of Eq.\ \ref{eq:sym_ham} (see App.\ \ref{app:diag}), we obtain the dependence of the energy levels on external flux as shown in Fig.\ \ref{fig:spectruma} \cite{Smith2016}. At $\varphi_\text{ext} = \pi$ (the dashed line in Fig.\ \ref{fig:spectruma}), the spectrum resembles a doubled harmonic oscillator with energy $\sqrt{16 x \epsilon_L \epsilon_C}$. Once $\varphi_\text{ext}$ deviates from $\pi$, half of the energy levels increase in energy linearly with slope $\sim \frac{32}{3} \epsilon_L$ \footnote{This term is simply the potential energy at $\varphi = \pi$ relative to that at $\varphi = 0$ (with $\phi_\text{ext} = 0$) found from Eq.\ \ref{eq:2m_ham}.}. The other half of the energy levels form a flux-independent harmonic ladder.

We can understand this level structure as that of two emergent modes (see also App.\ \ref{app:sector}). The first mode is flux-dependent and its excitations correspond to the number of magnetic flux vortices, or fluxons, enclosed by the inductive loop in Fig.\ \ref{fig:circuit}. In turn, the number of enclosed fluxons identically maps onto the magnitude and chirality of the circulating persistent current in the inductive loop. The second mode is flux-independent and its excitations correspond to quantized charge density oscillations, or plasmons, across the inductive loop/shunt capacitance in Fig.\ \ref{fig:circuit}. Each plasmon involves the two superinductances (energy $2\epsilon_L$ in parallel) and the shunt capacitance (energy $x \epsilon_C$), and hence has energy $\sqrt{16 x \epsilon_L \epsilon_C}$. Hereafter, we refer to these modes as the ``fluxon mode'' and the ``plasmon mode,'' respectively \footnote{We comment that the Aharonov-Bohm effect manifests itself in this circuit through the fluxon mode, in which charge carriers circle a pinned flux. Conversely, the Aharonov-Casher effect involves the plasmon mode, in which flux vortices circle a pinned charge.}. Additionally, we assign the labels $|m \blank\rangle$ to the lowest-energy states, where $m$ denotes the number of plasmons and $\blank$ (or $\Blank$) denotes the presence (or absence) of a fluxon excitation relative to the ground state \footnote{Here we restrict the Hilbert space to the two fluxon states with lowest energy.}. Note that the fluxon placeholder indices are defined by
\begin{equation*}
\blank/\Blank = \begin{cases} \currentL/\currentR, &\quad\text{for } \varphi_\text{ext}\!\!\!\!\! \mod{2\pi} < \pi \\
-/+, &\quad \text{for } \varphi_\text{ext} \!\!\!\!\! \mod{2\pi} = \pi \\
\currentR/\currentL, &\quad \text{for } \varphi_\text{ext} \!\!\!\!\!\mod{2\pi} > \pi \end{cases} ,
\end{equation*}
where $|m\pm\rangle = |m\rangle \otimes \frac{1}{\sqrt{2}} (|\currentR\rangle \pm |\currentL\rangle)$ (as in Sec.\ \ref{subsec:element}) and the index $\currentLR$ represents the persistent current direction. This labeling serves the purpose of assigning quantum numbers consistently for all external flux values (see the colors in Fig.\ \ref{fig:spectruma} \footnote{Note that the coloring changes for only the odd-numbered plasmon states, due to the fact that the eigenstate with even overall parity has lower energy for each plasmon state, and that the overall parity includes both a plasmon and a fluxon component.}), except the particular case $\varphi_\text{ext} \!\!\! \mod{2\pi} = 0$ that we do not focus on.

\subsection{Wavefunctions\label{subsec:wavefunctions}}

Calculated charge wavefunctions $\langle N | \psi \rangle$ and phase wavefunctions $\langle \varphi, \phi | \psi \rangle$, obtained from numerical diagonalization of Eq.\ \ref{eq:sym_ham}, are shown in Figs.\ \ref{fig:spectrumb}, \ref{fig:spectrumc} for the four lowest-energy eigenstates at $\varphi_\text{ext} = \pi$. Roughly, the phase wavefunctions are computed by projection of the $\theta$ coordinate and a Fourier transform to the $\varphi\phi$-plane, while the charge wavefunctions are computed by projection of the $\theta$ coordinate and constraint to the trajectory in Eq.\ \ref{eq:path} (see App.\ \ref{app:diag} for details).

The charge wavefunctions are grid states with Fock-state envelopes \cite{Gottesman2001,Bell2014}. For fluxon excitation index $+/-$, these grid states are superpositions of even/odd Cooper pair number states \footnote{The correspondence between Cooper pair number parity and the symmetry of the phase wavefunction can be seen from the $\pi$-periodicity/$\pi$-antiperiodicity of $|\!+\!/-\protect\rangle$ and $\protect\langle N | \psi \protect\rangle = \frac{1}{2\pi}\int_0^{2\pi} \text{d}\varphi \, \text{e}^{i N \varphi} \protect\langle \varphi | \psi \protect\rangle$  \cite{Devoret1997}.}. Additionally, $m$ corresponds to the order of the Fock state envelope. Note that a logical qubit encoded in $|0 + \rangle$ and $|0 - \rangle$ is protected from spurious transitions except those mediated by operators that flip Cooper pair parity. 

On the other hand, the phase wavefunctions are approximately Fock states localized within the potential energy wells (see Sec.\ \ref{subsec:semiclassical}) \cite{Earnest2018}. The fluxon index $+/-$ denotes whether the state $|m \pm\rangle$ is a symmetric (bonding) or antisymmetric (antibonding) superposition of states localized within opposite ridges of potential wells. These ridges correspond to persistent currents of opposite chirality, and hence also to the absence/presence of a fluxon in the inductive loop of the circuit \cite{Dempster2014}. In this picture, $m$ refers to the Fock order of the localized states. Finally, operators that flip Cooper pair parity correspond to odd functions of $\phi$ or functions of $\varphi$ with period an odd division of $2\pi$, which can be seen from Fig.\ \ref{fig:spectrumc} to mediate the transition $|m+\rangle \leftrightarrow |m-\rangle$

\section{Qubit\label{sec:qubit}}

In Sec.\ \ref{sec:circuit}, we analyzed the multi-mode Hamiltonian describing the superconducting circuit in Fig.\ \ref{fig:circuit}. Numerical diagonalization of this Hamiltonian showed the emergence of a linear plasmon mode and a nonlinear fluxon mode \footnote{The number of plasmon and fluxon modes form the rows and columns of a crude superconducting circuit periodic table. Charge-based qubits such as the transmon have one plasmon mode and no fluxon modes. Flux-based qubits such as the fluxonium have one fluxon mode and no plasmon modes. This circuit has two plasmon modes (one of which is neglected at low frequencies) and one fluxon mode. The larger resulting Hilbert space is the resource used to achieve protection.}. In this section, we consider the properties of the logical qubit formed by $\{|0+\rangle, |0-\rangle\}$, the two lowest-energy eigenstates at $\varphi_\text{ext} = \pi$, which generalizes to $\{|0\Blank\rangle, |0\blank\rangle\}$ away from $\varphi_\text{ext} = \pi$.

\subsection{Matrix elements\label{subsec:matrix}}

To better elucidate which types of operators can and cannot induce transitions between the two states of the qubit, we examine the relevant matrix elements corresponding to capacitive and inductive coupling. This discussion is particularly relevant to understanding the expected dominant loss mechanisms (Sec.\ \ref{subsec:T1}) and designing a measurement and control apparatus that does not directly couple to the qubit (Sec.\ \ref{subsec:control}).

For capacitive coupling, a generic voltage $V$ couples to the superconducting island of the circuit in Fig.\ \ref{fig:circuit} via a gate capacitance $C_\text{g}$ and will append the term
\begin{equation*}
H_\text{int} = \frac{C_\text{g}}{C_\text{shunt} + C_\text{g}} (2e\eta) V 
\end{equation*}
to the Hamiltonian in Eq.\ \ref{eq:sym_ham}, in addition to dressing the shunt capacitance. This voltage may be a degree of freedom of another mode in the embedding circuit, a noise source, or an ac drive. We therefore see that the susceptibility of undergoing a transition from the ground state, due to capacitive coupling to the qubit island, is directly related to the matrix element $\langle \psi | \eta | 0\Blank \rangle$.

For inductive coupling, a generic current $I$ couples to the circuit via a small inductance $L_\text{s}$ shared with the inductive loop, which adds the term
\begin{equation*}
H_\text{int} = \frac{L_\text{s}}{2L} \left(\phi_0 \phi \right) I 
\end{equation*}
to the Hamiltonian in Eq.\ \ref{eq:sym_ham}. Here, $\phi_0 = \hbar /2e$ is the reduced magnetic flux quantum and $L$ is the superinductance in each arm of the qubit (i.e.\ $\epsilon_L = \phi_0^2 / L$). Like the voltage source, this current may represent an internal or environmental degree of freedom. We see that the susceptibility of undergoing a transition from the ground state, due to inductive coupling to the inductive loop, is related to the matrix element $\langle \psi | \phi | 0\Blank \rangle$. 

\begin{figure}
\subfloat{%
\includegraphics{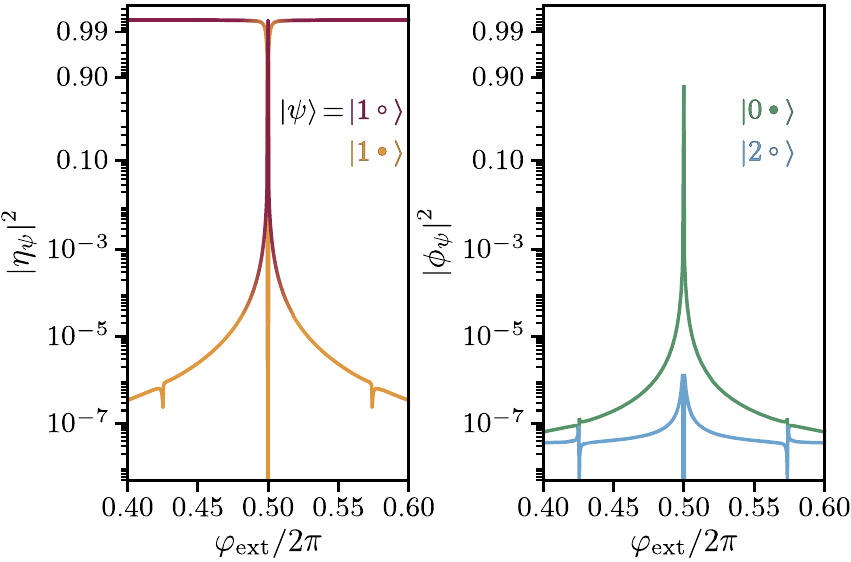}
\put(-245,157){(a)}
\put(-125,157){(b)}
\label{fig:matrixa}}
\subfloat{
\label{fig:matrixb}}
\caption{(a) Normalized charge matrix elements $|\eta_\psi|^2$ between the ground state $|0\Blank\rangle$ and the excited state $|\psi\rangle$, showing immunity of the qubit manifold to capacitive coupling. (b) Normalized phase matrix elements $|\phi_\psi|^2$ between states $|0\Blank\rangle$ and $|\psi\rangle$. This matrix element is near-unity between the two qubit states at $\varphi_\text{ext} = \pi$, meaning the qubit manifold is primarily susceptible to inductive loss.\label{fig:matrix}}
\end{figure}

Limiting the Hilbert space to the six lowest-energy eigenstates $\{ |m\pm\rangle : m = 0, 1, 2\}$, we numerically compute the normalized matrix elements
\begin{equation*}
|\mathcal{O}_\psi|^2 \equiv \frac{|\langle \psi | \mathcal{O} | 0\Blank \rangle|^2}{\langle 0\Blank | \mathcal{O}^\dagger \mathcal{O} | 0\Blank \rangle}
\end{equation*}
from the ground state $|0\Blank\rangle$ for the operators $\mathcal{O} = \eta, \phi$. Results are plotted in Fig.\ \ref{fig:matrix}. Note that $\sum_\psi |\mathcal{O}_\psi|^2 = 1$ and $|\mathcal{O}_\psi|^2 > 0$, so we may reasonably consider these as transition probabilities via $\mathcal{O}$. 

We see from Fig.\ \ref{fig:matrixa} that transitions mediated by capacitive coupling to the qubit island are only allowed from $|0\Blank\rangle$ to $|1\Blank\rangle$. These selection rules result from the decoupling of the even and odd Cooper pair number parity manifolds (see App.\ \ref{app:sector}). Most importantly, transitions between qubit states are forbidden, meaning capacitive coupling offers a promising ingredient for indirect qubit measurement and control (see Sec.\ \ref{subsec:control}). Conversely, inductive coupling to the inductive loop of the qubit permits transitions between $|0\Blank\rangle$ and $|0\blank\rangle$ in the vicinity of $\varphi_\text{ext} = \pi$, as shown in Fig.\ \ref{fig:matrixb}. This effect arises because the operator $\phi$ induces transitions between the Cooper pair parity manifolds, as can be seen from the Fourier series for Eq.\ \ref{eq:path}. As a consequence, we expect that relaxation of the qubit will be primarily due to inductive loss in the superinductances (see Sec.\ \ref{subsec:T1}).

\subsection{Disorder\label{subsec:disorder}}

A highly symmetric superconducting circuit is usually fragile in view of unavoidable fabrication imperfections \cite{Dempster2014}. The symmetry of our circuit involving the two inductive arms in Fig.\ \ref{fig:circuit} may be broken in three parameters: the Josephson energies of the junctions, the capacitances of the junctions, or the superinductances. To analyze these effects, we numerically diagonalize Eq.\ \ref{eq:sym_ham} and examine the energy splitting $\Delta E$ \footnote{The energy splitting $\Delta E$ is evaluated at $N_\text{g} = 0$.} as well as the charge dispersion $\epsilon = \max_{N_\text{g}} \Delta E - \min_{N_\text{g}} \Delta E$ of the $\{|0+\rangle,|0-\rangle\}$ manifold at $\varphi_\text{ext} = \pi$. A dimensionless quantity $\delta \in [0,1)$ is introduced to parameterize the extent of asymmetry in all three cases, and the $\delta$-dependence of the energies $\Delta E$ and $\epsilon$ is studied.

\subsubsection{Disorder in $\epsilon_J$\label{ssubsec:ej}}

\begin{figure}
\centering
\subfloat{%
\includegraphics{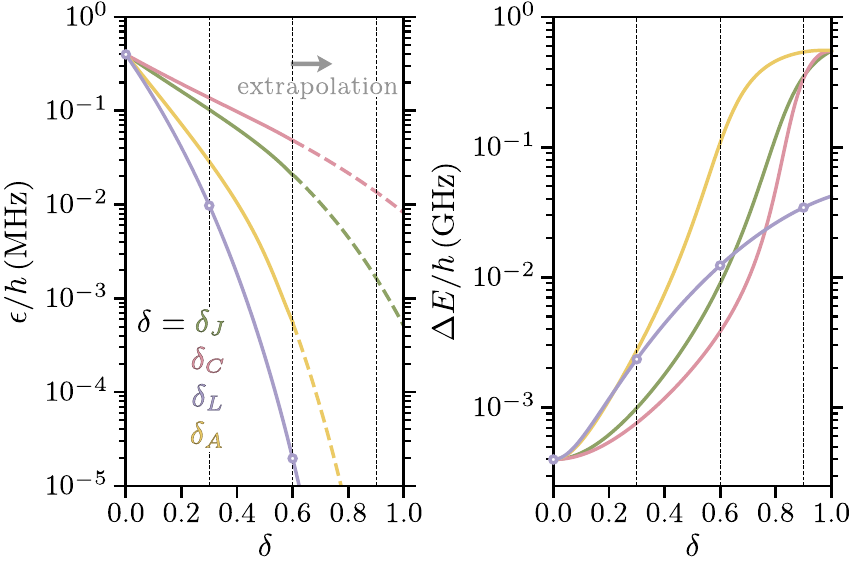}
\put(-245,159){(a)}
\put(-122,159){(b)}
\label{fig:disordera}}
\subfloat{
\label{fig:disorderb}}
\caption{(a) Plot of the charge dispersion $\epsilon$ of the qubit transition as a function of disorder parameters $\delta$. Here, $\delta_J$, $\delta_C$, and $\delta_L$ correspond to disorder in the characteristic energy scales $\epsilon_J$, $\epsilon_C$, and $\epsilon_L$, respectively. Disorder in the area of the two Josephson junctions is represented by $\delta_A$. (b) Plot of the energy splitting $\Delta E$ as a function of disorder parameters $\delta$. The dashed lines and circles indicate the values of $\delta_L$ used for inductive disorder in Sec.\ \ref{subsec:T1}, \ref{subsec:T2}.\label{fig:disorder}}
\end{figure}

We model disorder in the Josephson energies of the junctions by allowing the values of $\epsilon_J$ to deviate. We therefore set the left and right junction tunneling energies to $(1 \pm \delta_J)\epsilon_J$, respectively, where $\delta_J$ is the aforementioned asymmetry parameter. The Hamiltonian in Eq.\ \ref{eq:sym_ham} is perturbed by the term
\begin{equation*}
H^\prime = 2 \epsilon_J \delta_J \sin \varphi \sin \frac{\phi}{2}.
\end{equation*}
See Fig.\ \ref{fig:disorder} for a plot of $\Delta E$ and $\epsilon$ as a function of $\delta_J$. The important feature in these plots is that the charge dispersion decreases exponentially while the splitting increases exponentially with $\delta_J$ \footnote{The extrapolation of the charge dispersion as a function of the disorder beyond $\delta = 0.6$ is due to numerical instabilities inherent to low-energy quantities paired with disappearance of an efficient diagonalization basis for large disorder. This is particularly obvious in the vicinity of $\delta = 1$, at which point the number of modes of the circuit reduces from three to two.}. These features arise from the effective Hamiltonian in Eq.\ \ref{eq:2m_ham} being accompanied by $2\pi$-periodic terms in the presence of disorder. In this case of disorder in $\epsilon_J$, the approximations in Sec.\ \ref{subsec:semiclassical} lead to $H^\prime\approx H_\text{eff}^\prime$ with
%
%
\begin{equation*}
H^\prime_\text{eff} = - \frac{16}{3\pi} \epsilon_J \delta_J \left( \sin \varphi - \frac{1}{5} \sin 3 \varphi\right),
\end{equation*}
which evidently permits the tunneling of single Cooper pairs across the element. The resulting qubit retains characteristics of the symmetric circuit as well as the asymmetric/transmon-like circuit.

\subsubsection{Disorder in $\epsilon_C$\label{ssubsec:ec}}

Analogous to Sec.\ \ref{ssubsec:ej}, we set the left and right junction charging energies to $\epsilon_C/(1 \pm \delta_C)$, respectively. Aside from dressing the charging energy, the Hamiltonian in Eq.\ \ref{eq:sym_ham} inherits the term
\begin{equation*}
H^\prime = - 8 \epsilon_C \frac{\delta_C}{1 - \delta_C^2} n (N - N_\text{g} - \eta),
\end{equation*}
whose effect on the qubit manifold is plotted in Fig.\ \ref{fig:disorder}.

This form for capacitive disorder is assumed due to the fact that, if $\delta_J = \delta_C \equiv \delta_A$, then the product $\epsilon_J \epsilon_C$ is kept constant for both junctions under the effects of disorder. This corresponds to the physical case where the junction plasma frequencies are fixed by oxidation, but their areas differ due to fabrication imperfections. Here, the junction areas are $(1 \pm \delta_A) A$ because $A \propto \sqrt{\epsilon_J/\epsilon_C}$. The consequences of area disorder are plotted in Fig.\ \ref{fig:disorder}.

\subsubsection{Disorder in $\epsilon_L$\label{ssubsec:el}}

Following the same procedure, we set the left and right superinductance inductive energies to $\epsilon_L/(1 \pm \delta_L)$, respectively. This form is taken in order to fix the total linear inductance in the loop. Aside from dressing the inductive energy, the Hamiltonian in Eq.\ \ref{eq:sym_ham} is perturbed by
\begin{equation}
H^\prime = \epsilon_L \frac{\delta_L}{1 - \delta_L^2} (\phi - \varphi_\text{ext}) \theta. \label{eq:el}
\end{equation}
Note in Fig.\ \ref{fig:disorder} that the charge dispersion and energy splitting follow the same general trend for inductive disorder as for the other three. The key difference is that the charge dispersion decreases more quickly than for any other form of disorder. Oppositely, the splitting is initially the same as for area disorder, but the slope decreases in $\delta_L$. We conclude that disorder allows us to engineer a circuit with a sufficiently non-degenerate ground state manifold whose charge dispersion is largely suppressed. For reasons that will become clear in Sec.\ \ref{subsec:T2}, these features are extremely valuable for designing a qubit that is protected from dephasing.

\subsection{Relaxation\label{subsec:T1}}

\begin{table*}
\begin{ruledtabular}
\begin{tabular}{p{2.7cm} c D{?}{\,\times\,}{6.2} D{.}{.}{6.4} D{.}{.}{5.3} D{.}{.}{4.2} D{.}{.}{3.5}}
\T\B Loss channel & $\mathcal{O}$ & \multicolumn{1}{c}{Quality factor} & \multicolumn{4}{c}{$T_1$ (ms)} \\ \cline{4-7}
\T & & & \multicolumn{1}{c}{$\delta_L = 0.0$} & \multicolumn{1}{c}{$\delta_L = 0.3$} & \multicolumn{1}{c}{$\delta_L = 0.6$} &  \multicolumn{1}{c}{$\delta_L = 0.9$} \\ \hline
\T\B Capacitive & $2e N_i$ & Q_\text{cap} \sim 1 ? 10^6 \, \text{\cite{Wang2015}} & 780\,000 & 17\,000 & 1\,000 & 18 \\
\T\B Inductive & $\phi_0 \phi_i$ & Q_\text{ind} \sim 500 ? 10^6 \, \text{\cite{Pop2014}} & 0.61 & 0.79 & 1.1 & 1.4 \\
\T\B Purcell & $2e\eta$ & Q_\text{cap} \sim 1 ? 10^6 \, \text{\cite{Wang2015}} & \infty & 2\,500 & 380 & 470 \\
\T\B Quasiparticle & $2\phi_0 \sin \frac{\varphi_i}{2}$ & 1/x_\text{qp}^* \sim 0.3 ? 10^6 \, \text{\cite{Pop2014}} & \infty &  \infty & \infty & \infty \\ \hline\hline
\T\B Dephasing channel & $\lambda$ & \multicolumn{1}{c}{Spectral density amplitude} & \multicolumn{4}{c}{$T_\upphi$ (ms)} \\ \cline{4-7}
\T & & & \multicolumn{1}{c}{$\delta_L = 0.0$} & \multicolumn{1}{c}{$\delta_L = 0.3$} & \multicolumn{1}{c}{$\delta_L = 0.6$} & \multicolumn{1}{c}{$\delta_L = 0.9$} \\ \hline
\T\B Charge & $N_\text{g}$ & \sqrt{A_{N_\text{g}}}^* \sim 1 ? 10^{-4} \, \text{\cite{Zorin1996}} & 0.0037 & 0.15 & 74 & 3.3\times 10^6 \\
\T\B Flux & $\varphi_\text{ext}$ & \sqrt{A_{\varphi_\text{ext}}}/2\pi \sim 3 ? 10^{-6} \, \text{\cite{Quintana2017}} & 0.022 & 0.13 & 0.67 & 1.8 \\
\T\B Shot & $n_\text{p}$ & n_\text{th}/Q_\text{cap} \sim 1 ? 10^{-7} \, \text{\cite{Wang2015}} & 4.6 & 4.8 & 5.3 & 8.7 \\
\T Critical current & $\epsilon_J$ & \sqrt{A_{\epsilon_J}}/\epsilon_J \sim 5 ? 10^{-7} \, \text{\cite{VanHarlingen2004}} & 210 & 40 & 8.2 & 2.9 \\
\end{tabular}
\end{ruledtabular}
\caption{Expected relaxation times $T_1$ and pure dephasing times $T_\upphi$ at $\varphi_\text{ext} = \pi$ through various channels. The operators $\mathcal{O}$ coupling to the bath and the noisy parameters $\lambda$ are listed in addition to the associated quality factors and spectral density amplitudes, respectively. Relaxation and dephasing times are shown for varying cases of inductive disorder: $\delta_L = 0.0, 0.3, 0.6, 0.9$ (see also Fig.\ \ref{fig:disorder}). The operators $N_i$ depict the number of Cooper pairs having tunneled across the two Josephson junctions, i.e.\ $N_i = n \pm \frac{1}{2} (N - \eta)$ Entries that read ``$\infty$'' represent numerical infinity. *These values are reported for clarity but have no bearing on the estimates shown (see Secs.\ \ref{subsec:T1}, \ref{subsec:T2}). \label{tab:coherence}}
\end{table*}


We model loss due to an arbitrary channel using Fermi's Golden Rule. This gives us the relaxation rate, including both emission and absorption, of
\begin{equation}
\frac{1}{T_1} = \frac{1}{\hbar^2} | \langle 0 \! + \! | \mathcal{O} | 0- \rangle |^2 \left[ S_\mathcal{EE} (\Delta \omega) + S_\mathcal{EE} (- \Delta \omega) \right]. \label{eq:T1}
\end{equation}
In this expression, $\mathcal{O}$ is the operator of the circuit coupling to a noisy bath variable $\mathcal{E}(t)$, the noise spectral density of which is given by $S_\mathcal{EE} (\omega)$. Note that $\Delta E = \hbar \,\Delta \omega$. In this subsection, we calculate the relaxation rates for the expected dominant loss mechanisms of the qubit based on numerical diagonalization of the full Hamiltonian in Eq.\ \ref{eq:sym_ham}. We perform this calculation for various degrees of inductive disorder $\delta_L$ (see Fig.\ \ref{fig:disorder}).

We consider four possible loss channels for the qubit: capacitive loss, inductive loss, Purcell loss, and quasiparticle tunneling \cite{Pop2014}. Capacitive loss involves dielectric dissipation in the Josephson junction capacitances. In this case, we have $\mathcal{O} = 2e N_i = 2e \big[n \pm \frac{1}{2} (N - \eta)\big]$ for the charge across the $i$-th junction and $\mathcal{E} = V$ for a bath voltage with
\begin{equation}
S_{VV}(\omega) + S_{VV}(-\omega) = \frac{2\hbar}{C_J \mathcal{Q}_\text{cap}(\omega)} \coth \frac{\hbar|\omega|}{2k_\text{B} T}. \label{eq:svv}
\end{equation}
Here, $C_J = e^2 / 2\epsilon_C$ is the junction capacitance, $T$ is the temperature, and $\mathcal{Q}_\text{cap}(\omega)$ is a frequency-dependent quality factor \cite{Braginsky1987} with nominal value $Q_\text{cap} \sim 1 \times10^6$ \cite{Wang2015} (see App.\ \ref{app:estimates} for more details).

Depending on the specific implementation, inductive loss may occur within the superinductances via quasiparticle tunneling \cite{Pop2014}. This situation can be modeled by taking $\mathcal{O} = \phi_0 \phi_i$ for the flux across the $i$-th superinductance and $\mathcal{E} = I$ for a bath current with
\begin{equation*}
S_{II}(\omega) + S_{II}(-\omega) = \frac{2\hbar}{L_i \mathcal{Q}_\text{ind}(\omega)}\coth \frac{\hbar |\omega|}{2k_\text{B} T}.
\end{equation*}
In this expression, $L_i = (1 \pm \delta_L) L$ is the $i$-th superinductance and $\mathcal{Q}_\text{ind}(\omega)$ is a frequency-dependent quality factor with nominal value $Q_\text{ind} \sim 500 \times 10^6$ \cite{Pop2014}. As shown in App.\ \ref{app:estimates}, this frequency dependence is included to extrapolate to small qubit transition frequencies.

Loss due to the Purcell effect should mainly arise from coupling of the qubit to the plasmon mode, which we model as dielectric loss in the shunt capacitance. As such, we have $\mathcal{O} = 2e \eta$ and $\mathcal{E} = V$ for a bath voltage with the same noise spectral density as Eq.\ \ref{eq:svv}, but with $C_J$ replaced by $C_\text{shunt}$.

Finally, quasiparticle tunneling is expected to occur across either Josephson junction and contribute to qubit relaxation \cite{Pop2014}. In this case, we have $\mathcal{O} = 2\phi_0 \sin \frac{\varphi_i}{2}$ for the $i$-th junction and the noise spectral density
\begin{equation*}
S_\text{qp}(\omega) + S_\text{qp}(-\omega) = 2 \hbar \omega \, \text{Re} \, Y_\text{qp}(\omega) \coth \frac{\hbar\omega}{2k_\text{B} T},
\end{equation*}
with $\text{Re}\, Y_\text{qp}(\omega)$ being the dissipative part of the Josephson junction admittance \cite{Catelani2011} (see App.\ \ref{app:estimates} for the explicit form). Note that this admittance scales linearly in both $\epsilon_J$ and the quasiparticle density $x_\text{qp}$ defined relative to that of Cooper pairs.

The calculated relaxation rates and the corresponding components of Eq.\ \ref{eq:T1} are shown in Tab.\ \ref{tab:coherence} for all four loss channels at $\varphi_\text{ext} = \pi$. For both capacitive and inductive loss, we note that measurements have not yet been performed using qubits with extremely small energy splittings, and the calculation presented here relies on an extrapolation to such frequencies. A key feature is the complete absence of quasiparticle loss, as in the fluxonium qubit \cite{Pop2014}. Additionally, we see that the asymmetric qubit is marginally less susceptible to inductive loss than the symmetric qubit. This improvement comes at the cost of the susceptibility to capacitive and Purcell loss. However, we emphasize that the lifetimes shown in Tab.\ \ref{tab:coherence} are conservative estimates that demonstrate $T_1 \gtrsim 1$ ms, which is at least competitive with state-of-the-art qubit implementations \cite{Rigetti2012,Barends2013,Pop2014,Bell2014,Yan2016,Groszkowski2018}.

\subsection{Pure dephasing\label{subsec:T2}}

In this subsection, we examine the dependence of the qubit transition energy $\Delta E$ on various system parameters $\lambda$ corresponding to different dephasing mechanisms. We estimate the dephasing times due to offset charge noise, external flux noise, photon shot noise in the plasmon mode, and critical current noise. The noise spectral densities are all assumed to be $1/f$ and hence given by $S_{\lambda\lambda}(\omega) = 2\pi A_\lambda/|\omega|$, where $\sqrt{A_\lambda}$ is the amplitude, with the sole exception of shot noise whose spectral density is assumed to be Lorentzian.

\subsubsection{Charge noise\label{ssubsec:charge}}

In the case of perfect symmetry, the charge dispersion $\epsilon$ is identically mapped to $\Delta E$. As a consequence, resilience to dephasing from offset charge noise demands a high degree of degeneracy, making experimental implementation difficult. Fortunately, this issue can be avoided by introducing inductive disorder into the system (see Sec.\ \ref{ssubsec:el}). In the slow-varying limit for charge noise, the pure dephasing rate is bounded by
\begin{equation*}
\frac{1}{T_\upphi} = \frac{\pi}{(2\text{e})^2} \epsilon/\hbar,
\end{equation*}
where $\text{e}$ is Euler's number \cite{Koch2007.1}. In Tab.\ \ref{tab:coherence}, we see that this yields a strict bound on the decoherence time in the case of perfect symmetry, which is greatly alleviated in the presence of inductive disorder. Note that this estimate does not explicitly involve $\sqrt{A_{N_\text{g}}}$ because the slow-varying limit has been taken where the offset charge assumes a random value for each measurement, but does not fluctuate within a given measurement.

\subsubsection{Flux noise\label{ssubsec:flux}}

At $\varphi_\text{ext} = \pi$, the qubit manifold is only susceptible to external flux noise to second order. However, one can easily obtain \footnote{For example, this can done using the double-well Hamiltonian $H = \frac{1}{2} (\Delta E) \sigma_x + \frac{1}{2} \pi \epsilon_L (\varphi_\text{ext} - \pi) \sigma_z$ for $\varphi_\text{ext}$ in the vicinity of $\pi$, where the two spin states correspond to $\phi = 0, \pi$. One then has the flux-dependent splitting $\delta E = \sqrt{(\Delta E)^2 + \pi^2 \epsilon_L^2 (\varphi_\text{ext} - \pi)^2}$ and hence $\partial^2 (\delta E)/\partial \varphi_\text{ext}^2 = (\pi \epsilon_L)^2  / \Delta E$ (at $\varphi_\text{ext} = \pi$).} the relation $\partial^2 \Delta E / \partial \varphi_\text{ext}^2 \propto 1/\Delta E$ (at $\varphi_\text{ext} = \pi$), which shows that insensitivity to flux-noise-induced dephasing requires sufficiently weak degeneracy, at odds with the requirement for resilience to charge noise. Once more, introducing inductive disorder avoids this issue. Neglecting components that depend on details of the implementation, the pure dephasing rate \cite{Cottet2002.2} is bounded by
\begin{equation*}
\frac{1}{T_\upphi} = A_{\varphi_\text{ext}} \bigg| \frac{\partial^2 \Delta E}{\partial \varphi_\text{ext}^2} \bigg|,
\end{equation*}
where $\sqrt{A_{\varphi_\text{ext}}}/2\pi \sim 3 \times 10^{-6}$ is the amplitude of the noise spectral density \cite{Quintana2017,Kou2017} for $\varphi_\text{ext}$. As in the case of charge noise, this places a stringent limit on the decoherence time of the qubit in the absence of inductive disorder, as we see in Tab.\ \ref{tab:coherence}.

\subsubsection{Shot noise\label{ssubsec:shot}}

We expect the dominant contribution to dephasing due to thermal photons to arise from coupling to the plasmon mode. This mode has an angular frequency $\omega_\text{p} \approx \sqrt{16x\epsilon_C \epsilon_L}/\hbar$ and an occupation $n_\text{p}$ with mean given by $n_\text{th} = 1/(\text{e}^{\hbar \omega_\text{p}/k_\text{B} T} - 1)$. To this end, we place a bound on the pure dephasing time $T_\upphi$ using the expression \cite{Rigetti2012}
\begin{equation*}
\frac{1}{T_\upphi} = n_\text{th} \kappa \frac{\chi^2}{\chi^2 + \kappa^2},
\end{equation*}
where $\chi$ is the dispersive shift of the qubit on the plasmon mode and $\kappa = \omega_\text{p} / \mathcal{Q}_\text{cap}(\omega_\text{p})$ is the linewidth of the plasmon mode. Note that, as in Sec.\ \ref{subsec:T1}, we use the frequency-dependent dielectric quality factor $\mathcal{Q}_\text{cap}(\omega)$ with nominal value $Q_\text{cap} \sim 1 \times 10^6$ \cite{Wang2015}. In all cases, this form of noise is not expected to limit the decoherence time of the qubit (see Tab.\ \ref{tab:coherence}).

\subsubsection{Critical current noise\label{ssubsec:current}}

The final dephasing mechanism we investigate is critical current noise, or fluctuations in the Josephson energy $\epsilon_J$. As in the case of flux noise, we discard factors that are sensitive to experimental details and take the dephasing rate to be bounded by \cite{Cottet2002.2}
\begin{equation*}
\frac{1}{T_\upphi} = \sqrt{A_{\epsilon_J}} \left| \frac{\partial \Delta E}{\partial \epsilon_J} \right|,
\end{equation*}
where $\sqrt{A_{\epsilon_J}} \sim 5 \times 10^{-7} \epsilon_J$ is the amplitude of the noise spectral density \cite{VanHarlingen2004} for $\epsilon_J$. As in the case of photon shot noise, we do not expect critical current noise to place a strict bound on the decoherence time (see Tab.\ \ref{tab:coherence}).

At this point, it is clear that inductive asymmetry constitutes a necessary ingredient in the protection of this qubit. In light of Eq.\ \ref{eq:el}, we attribute this to the resulting hybridization of the $\phi$ and $\theta$ modes, which are not directly coupled in the symmetric case (see Eq.\ \ref{eq:sym_ham}). The fluxon transition between the qubit states inherits some character of the plasmon transition, thereby breaking the correspondence between $\epsilon$ and $\Delta E$ and reducing the flux matrix element $\langle 0\!-\! | \phi | 0+ \rangle$.

\section{Discussion\label{sec:discussion}}

\subsection{Protection\label{subsec:protection}}

We have proposed a superconducting circuit whose ground state manifold can be made protected, at the Hamiltonian level, from multiple common sources of noise. Here we describe some of the other strategies for achieving protection. The simplest manifestations involve improving quality factors associated with coupling to different thermal baths \cite{Martinis2005,Houck2008,Barends2013} and reducing noise spectral densities for different Hamiltonian parameters \cite{Paik2011,Rigetti2012,Kumar2016}. At a higher level, a popular approach to suppressing qubit relaxation has been to localize wavefunctions in disparate regions of phase space to lessen transition matrix elements \cite{Pop2014,Bell2014,Earnest2018,Lin2018}. On the other hand, delocalization of the same wavefunctions has been shown to mitigate dephasing effects by reducing qubit sensitivity to Hamiltonian parameters \cite{Wal2000,Vion2002,Koch2007.1,Manucharyan2009,Steffen2010,Yan2016}. Superconducting circuits with multiple degrees of freedom whose qubit wavefunctions are both localized and delocalized combine both of these approaches. In these circuits, quantum information is diffused among constituent local degrees of freedom, providing protection from local perturbations. The many-body limit promises topological protection \cite{Doucot2002,Ioffe2002.1,Doucot2003,Kitaev2006}, in which global operators are necessary to manipulate logical qubits, but the few-body case described here offers an experimentally realistic approximation \cite{Cottet2002.1,Gladchenko2009,Brooks2013,Bell2014,Dempster2014,Kou2017,Groszkowski2018}. In the following subsection, we outline the key differences between our circuit and similar proposals.

\subsection{Comparison to other protected qubits\label{subsec:compare}}

\subsubsection{Heavy fluxonium\label{ssubsec:heavy}}

The heavy fluxonium \cite{Earnest2018,Lin2018}, in which a conventional fluxonium qubit is shunted by a large capacitance, shares many similar features with the circuit in Fig.\ \ref{fig:circuit}. In fact, our circuit roughly reduces to that of the conventional fluxonium in the extremely asymmetric limit of $\delta_L \rightarrow 1$, and the eigenstates of both circuits are similarly represented in terms of persistent currents. The two main differences are that our circuit contains an additional small Josephson junction and that the shunt capacitance is placed across the junction along with a subset of the array junctions. An analogy can be drawn between $\delta_L$ in our circuit and the shunt capacitance in the heavy fluxonium, because they both suppress the qubit transition frequency. Physically, the qubit eigenstates differ when these analogous parameters fall to small values. While the eigenstates collapse to Cooper pair number parity states in our circuit, this is not true for the heavy fluxonium because the variable $\varphi$ is not compact.

\subsubsection{Rhombus qubit\label{ssubsec:rhombus}}

The circuit in Fig.\ \ref{fig:circuit} bears resemblance to the rhombus \cite{Blatter2001}, with two central differences. First, in the rhombus, the superinductances are replaced by single Josephson junctions, or arrays of a few larger junctions \cite{Doucot2012,Bell2014}. This changes the parameter regime of the circuit from $\epsilon_L \ll \epsilon_J$ to $\epsilon_L \sim \epsilon_J$, decreasing the amplitude of the $\cos 2\varphi$ term in the Hamiltonian. Second, the shunt capacitance is replaced by a gate capacitance to a voltage source \cite{Bell2014}. This is akin to substituting an electrostatic gate for a shunt capacitance to obtain the Cooper pair box \cite{Vion2002} from the transmon \cite{Koch2007.1}; overall suppression of the charge dispersion is traded for the ability to bias the circuit at its charge sweet spot.

Finally, the rhombus by itself is not designed to be a protected qubit. Rather, when multiple rhombi are arranged into a one-dimensional chain (or a two-dimensional fabric), the ground states are eigenstates of a nonlocal operator, which provides topological protection \cite{Doucot2002,Ioffe2002.1,Doucot2003,Doucot2012}. On one hand, our qubit does not require such scaling to achieve protection. On the other hand, the protection we predict is inherently susceptible to local perturbations (see Secs.\ \ref{subsec:T1}, \ref{subsec:T2}).

\subsubsection{0-$\pi$ qubit\label{ssubsec:zeropi}}

The 0-$\pi$ qubit \cite{Brooks2013} also has a similar superconducting circuit to that in Fig.\ \ref{fig:circuit}, but with three essential distinctions. First, the pairs of superinductances on each arm of the circuit are combined, altering the embedding capacitance matrix, in particular the capacitances to ground. Second, there is the addition of a second large capacitance shunting the inductive loop between its two horizontally oriented nodes. When this second capacitance is precisely $C_\text{shunt}$, this permits the exact decoupling of the $\theta$ mode from the $\varphi$ mode in Eq.\ \ref{eq:sym_ham}. Third, the 0-$\pi$ qubit is operated in a parameter regime where $\epsilon_L \sim 0.01 \epsilon_J$, as opposed to our circuit, where $\epsilon_L \sim 0.1 \epsilon_J$ \cite{Dempster2014}. This additional order of magnitude in the superinductance may prove marginally less accessible experimentally \cite{Manucharyan2009,Masluk2012}.

Notably, the inductive loop in the 0-$\pi$ qubit is threaded with $\varphi_\text{ext} \approx 0$ at its working point instead of $\varphi_\text{ext} = \pi$. This leads to a substantial change in the physics of the ground state manifold; $|0+\rangle$ and $|0-\rangle$ are approximately localized in distinct potential wells \cite{Dempster2014}. In our case, these states are approximately the symmetric and antisymmetric superpositions of the localized wavefunctions, which are themselves localized in distinct Cooper pair number parities.

\subsection{Readout and control\label{subsec:control}}

Protected qubits face the serious obstacle of realizing state manipulation and measurement while remaining sufficiently isolated from their environments to preserve their coherence. We envision performing readout and control using an ancillary mode structure, which enables cascaded dispersive readout and Raman indirect transitions, as outlined below.

For readout, we aim to exploit the sizable native dispersive shift of the qubit on the plasmon mode, $\chi/2\pi \sim -20\,\text{MHz}$ (see Sec.\ \ref{ssubsec:shot}). Unfortunately, the small anharmonicity, which is at most of order $10\,\text{MHz}$, of the plasmon mode makes readout of the plasmon mode using a linear mode (e.g.\ a microwave cavity) difficult. As a remedy, we propose introducing an ancillary anharmonic mode by which to measure the plasmon state. In this scheme, a dispersive interaction will be mediated by the ancillary mode between the readout and plasmon modes, thereby enabling dispersive measurement with two readout tones \cite{Kirchmair2013}.

The above-mentioned ancillary anharmonic mode will also be useful for control of the protected qubit. As shown in Sec.\ \ref{subsec:matrix}, direct transitions mediated by capacitive coupling to the qubit superconducting island are strictly forbidden in the symmetric case, and weakly forbidden in the asymmetric case. Note that this coupling method is chosen to minimize the contributions to decoherence resulting from the introduction of the readout/control circuit. For manipulation, we therefore propose transitioning through the $|1+\rangle$ or $|1-\rangle$ state, as in a conventional $\Lambda$ system. An additional complication is that, even in the presence of inductive asymmetry, the joint fluxon-plasmon transitions $|0+\rangle \leftrightarrow |1-\rangle$ and $|0-\rangle \leftrightarrow |1+\rangle$ are forbidden.

Using an ancillary mode based on the Superconducting Nonlinear Asymmetric Inductive eLement (SNAIL) \cite{Frattini2017}, it is possible to engineer selection rules where the joint fluxon-plasmon transitions are weakly permitted while the qubit transition is left unaffected, as has recently been demonstrated in a fluxonium qubit \cite{Vool2018}. This stems from the feature of the SNAIL that allows, at specific flux biases, third-order nonlinearities without fourth-order nonlinearities. If we parameterize the broken selection rule using $y \in [0,1)$, the matrix element of $|0+\rangle \leftrightarrow |1-\rangle$ relative to $|0+\rangle \leftrightarrow |1+\rangle$, then the gate speed via sequential direct transitions and a stimulated Raman transition is estimated to be $(1 + y^{-1}) t_\mathrm{gate}$ and $2 \Delta y^{-1} t_\mathrm{gate}^2$, respectively \cite{Steck2019}. Here, $\Delta \sim 100\,\text{MHz}$ is the detuning of the Raman drive from the intermediate state and $t_\mathrm{gate} \sim 10\,\text{ns}$ is the $\pi$-pulse time \cite{Chow2010}. This also introduces an additional Purcell loss channel $|0-\rangle \rightarrow |1+\rangle \rightarrow |0+\rangle$ (and vice versa), which corresponds to the relaxation time $T_1 \sim y^{-2} \mathcal{Q}_\text{cap}(\omega_\mathrm{p})/\omega_\mathrm{p}$. We therefore expect gate speeds to be slowed by a factor of $y^{-1} < 10$ while lifetimes are kept in the millisecond range.

\section{Conclusion\label{sec:conclusion}}

In summary, we have designed a few-body superconducting circuit in which the charge carriers are well-approximated by pairs of Cooper pairs at a particular bias point. The Josephson tunneling element that supports these charge carriers is characterized by a $\cos 2\varphi$ term in the Hamiltonian, whose emergence we have shown analytically. Our numerical simulations supplement these arguments and demonstrate protection against a variety of common relaxation and dephasing sources. We find that this protection is substantially enhanced in the presence of disorder. Finally, we compared our circuit to similar proposals and offered our perspectives on readout and control.

As a final remark, we comment that engineering a circuit whose potential energy is dominated by a $\cos 2\varphi$ term opens the door to more exotic designs with potential energies of the form $\cos \mu \varphi$, with $\mu \in \mathbb{N}$, which could be obtained by introducing additional loops in the circuit. These could be tremendously valuable for quantum simulation, realizing nearly degenerate ground state manifolds with greater multiplicities, or performing degeneracy-preserving measurements of photon number parity \cite{Cohen2017}.

\begin{acknowledgments}
We thank J.\ Cohen, S.\ M. Girvin, L.\ I. Glazman, M.\ Mirrahimi, R.\ J.\ Schoelkopf, K.\ Serniak, and S.\ Shankar for their insights. This work was supported by the Army Research Office Grant No.\ W911NF-14-1-0011 and W911NF-18-1-0212. W.C.S. was supported by DoDMURI award No. FP057123-C. Facilities use was supported by the Yale Institute of Nanoscience and Quantum Engineering under National Science Foundation Grant No. MRS1119826.
\end{acknowledgments}

%
%
%
%
%

\appendix
\section{Mathieu equation\label{app:mathieu}}

The time-independent Schr\"{o}dinger equation for Eq.\ \ref{eq:simple_ham} assumes the form
\begin{equation*}
4 E_C \left(i \frac{\partial}{\partial \varphi} + N_\text{g}\right)^2 \psi - E_J \cos 2 \varphi \, \psi = E \psi.
\end{equation*}
Using the trial function $g \equiv \text{e}^{-i N_\text{g} \varphi} \psi$, this equation becomes
\begin{equation*}
g^{\prime\prime} + \left(\frac{E_J}{4 E_C} \cos 2\varphi + \frac{E}{4 E_C} \right) g = 0,
\end{equation*}
which is the Mathieu equation \cite{Meixner1980,Koch2007.1}. The $k$-th eigenenergy is given by
\begin{equation*}
E_k(N_\text{g}) = 4 E_C a_{N_\text{g} + \ell(k,N_\text{g})} \left( - \frac{E_J}{8 E_C}\right),
\end{equation*}
where $a$ denotes the Mathieu characteristic value and $\ell$ is an integer that sorts the eigenvalues. The crucial difference between this solution and that for the Cooper pair box is that the characteristic exponent $2 (N_\text{g} + \ell)$ is replaced by $N_\text{g} + \ell$ \cite{Koch2007.1}. This gives rise to the nearly degenerate ground states in the Hamiltonian in Eq.\ \ref{eq:simple_ham}. In particular, for $E_J \gg E_C$, we can use the asymptotic form of the Mathieu characteristic values to find
\begin{equation*}
E_k(N_\text{g}) \approx E_k (1/2) - \frac{1}{2} \epsilon_k \cos (\pi N_\text{g})
\end{equation*}
with $\epsilon_k$ being the charge dispersion of the $k$-th level. Importantly, for even values of $k$, we have $E_k = E_{k+1} \approx (\sqrt{32 E_J E_C} - 2 E_C) k$ (neglecting constants and nonlinear terms) and $\epsilon_k = - \epsilon_{k+1}$. Additionally, the expression for the ground state splitting in Sec.\ \ref{subsec:element} is justified by
\begin{equation*}
\epsilon_k \approx (-1)^k \frac{4^{k+2}}{k!} E_C \sqrt{\frac{2}{\pi}} \left(\frac{2E_J}{E_C}\right)^\frac{2k+3}{4} \text{e}^{-\sqrt{2E_J/E_C}},
\end{equation*}
and the fact that $\Delta E = E_1 - E_0 = \epsilon_0 \cos (\pi N_\text{g})$.

\section{Instanton treatment\label{app:instanton}}

The average tunneling trajectory between potential minima for the Hamiltonian in Eq.\ \ref{eq:sym_ham} can be found by examining an equivalent classical problem. This is done by inverting the potential and solving the classical equations of motion for a particle that starts at one maxima and ends at the other, each with zero kinetic energy \cite{Matveev2002}. Consequently, this classical trajectory requires an infinite amount of time. The corresponding differential equations,
\begin{align*}
\frac{\hbar^2}{8 \epsilon_C} \biggl[ 2 \ddot{\varphi} + \frac{1}{x}( \ddot{\varphi} + \ddot{\theta}) \biggr] &= 2 \epsilon_J \sin\varphi \cos\frac{\phi}{2} \\
\frac{\hbar^2}{16 \epsilon_C} \ddot{\phi} &= \frac{1}{2}\epsilon_L (\phi - \varphi_\text{ext}) + \epsilon_J \cos\varphi \sin\frac{\phi}{2} \\
\frac{\hbar^2}{8 x \epsilon_C} (\ddot{\varphi} + \ddot{\theta}) &= 2 \epsilon_L \theta,
\end{align*}
are solved numerically. Carrying out the procedure in Sec.\ \ref{subsec:semiclassical}, but retaining terms through the fourth harmonic and $O(z^2)$, results in the extended version of the effective Hamiltonian
%
%
\begin{align*}
H_\text{eff} &= 4 \epsilon_C \left\{ \frac{1}{2} \left[ 1 + \frac{1}{(1 + z)^2}\right]^{-1} (N - N_\text{g} - \eta)^2 + x \eta^2 \right\} \\
&\hspace{0cm} + \epsilon_L \theta^2 - \epsilon_L \left( \frac{16}{3\pi} - \frac{56}{9\pi}z\right) (\pi - \phi_\text{ext}) \cos \varphi\\
&\hspace{0cm} - \epsilon_J\left\{1 - \frac{5}{4}z + \frac{1}{48}[81 -2\pi^2 - 6(\pi-\phi_\text{ext})^2]z^2\right\}\cos 2\varphi\\
&\hspace{0cm} + \epsilon_L \left( \frac{16}{45\pi} - \frac{88}{75\pi} z\right)(\pi - \phi_\text{ext}) \cos 3\varphi\\
&\hspace{0cm} - \epsilon_L \left(\frac{1}{12} - \frac{17}{72} z\right) \cos 4\varphi,
\end{align*}
which clearly reduces to Eq.\ \ref{eq:2m_ham}. At $\varphi_\text{ext} = \pi$, we see that the dominant correction term is of the form $\cos 4 \varphi$. Moreover, we observe that the odd Fourier terms exactly vanish for the symmetric circuit at this flux bias, while the coefficients for the even Fourier terms decay roughly in powers of $z$.

\section{Numerical diagonalization\label{app:diag}}

For numerical diagonalization of either Eq.\ \ref{eq:sym_ham} or Eq.\ \ref{eq:2m_ham}, we employ a charge basis for the $\varphi$ mode in order to efficiently capture the dependence of the Hamiltonian on $N_\text{g}$ \cite{Devoret1997}. This is due to the exact periodicity of the Hamiltonian in $\varphi$, which inhibits the use of a harmonic oscillator basis to capture the multi-well physics apparent in Fig.\ \ref{fig:instanton}. For the $\theta$ and $\phi$ modes, we use harmonic oscillator bases. Explicitly, the full Hamiltonian in Eq.\ \ref{eq:sym_ham} can be written as
\begin{align}
H &= \sqrt{8 \epsilon_L \epsilon_C} \, a^\dagger a + \sqrt{16 x \epsilon_L \epsilon_C} \, b^\dagger b \nonumber \\
&\hspace{0.5cm} + 2\epsilon_C \big[N - N_\text{g} - i \eta_\text{zpf} (b^\dagger - b) \big]^2 \nonumber \\
&\hspace{0.5cm} - 2\epsilon_J \cos \varphi \cos \Big[ \tfrac{1}{2} \phi_\text{zpf} (a^\dagger + a) + \tfrac{1}{2} \varphi_\text{ext} \Big] \label{eq:num_ham}
\end{align}
after discarding constants, where $a^\dagger/a$ and $b^\dagger/b$ are bosonic creation/annihilation operators for the $\phi$ and $\theta$ modes, respectively. Additionally, the zero point fluctuation amplitudes $\phi_\text{zpf} = \big(\frac{8\epsilon_C}{\epsilon_L}\big)^{1/4}$ and $\eta_\text{zpf} = \frac{1}{2} \big(\frac{\epsilon_L}{x\epsilon_C}\big)^{1/4}$ have been introduced for convenience. The diagonalization basis is then
\begin{equation*}
\{ |N p q \rangle  : |N| \leq N_0, p \leq p_0, q \leq q_0\}
\end{equation*}
with $|N\rangle$ being the Cooper pair number eigenstates of the operator $N$ and $|p\rangle / |q\rangle$ being the photon number eigenstates of $a^\dagger a/b^\dagger b$. For numerical convergence, the dimensions $N_0 \gtrsim 7$, $p_0 \gtrsim 7$, and $q_0 \gtrsim 30$ are used, depending on the desired truncation accuracy. The remaining operators are written as
\begin{align*}
N - N_\text{g} &= \sum_{N = -N_0}^{N_0} (N - N_\text{g}) |N\rangle \langle N| \\
\cos \varphi &= \frac{1}{2} \sum_{N = -N_0}^{N_0} \Big( |N\rangle \langle N+1| + |N+1\rangle \langle N|\Big)
\end{align*}
in this basis, while the final cosine term in Eq.\ \ref{eq:num_ham} has an exact matrix expression \cite{Smith2016}, which we do not repeat here. Although the above discussion applies to the case of perfect symmetry, the discussion in Sec.\ \ref{subsec:disorder} allows a straightforward extension to the disordered case. Finally, we comment that, despite the high dimensionality of the matrix in Eq.\ \ref{eq:num_ham}, diagonalization can be made efficient by taking advantage of its sparsity.

Now we detail the method used to compute the wavefunctions $\langle N | \psi\rangle$ and $\langle \varphi, \phi | \psi\rangle$ (up to an overall phase) for $|\psi\rangle = |m \pm\rangle$ plotted in Figs.\ \ref{fig:spectrumb}, \ref{fig:spectrumc}. First, the coordinate space wavefunction $\langle \varphi, \phi, \theta | \psi\rangle$ is constructed using $\langle \varphi | N \rangle = \text{e}^{-i N \varphi}$ \cite{Devoret1997} and the expressions for $\langle \phi | p\rangle$ and $\langle \theta | q\rangle$ as harmonic oscillator wavefunctions:
\begin{equation*}
\langle \varphi, \phi, \theta | \psi\rangle = \sum_{N = -N_0}^{N_0} \sum_{p = 0}^{p_0} \sum_{q = 0}^{q_0} c_{Npq} \langle \varphi | N\rangle \langle \phi | p \rangle \langle \theta | q \rangle,
\end{equation*}
where the complex coefficients $c_{Npq}$ are obtained by diagonalization (and consistent choice of overall phase). Then, the $\theta$-dependence is removed by projection onto $\theta = 0$ and normalization. This amounts to computing
\begin{equation*}
\langle \varphi, \phi | \psi \rangle = \frac{\langle \varphi, \phi, 0 | \psi\rangle}{\sqrt{\int_0^{2\pi} \mathrm{d}\varphi \int_{-\infty}^\infty \mathrm{d}\phi \, | \langle \varphi, \phi, 0 | \psi\rangle|^2}},
\end{equation*}
which is plotted in Fig.\ \ref{fig:spectrumc}. Next, the $\phi$-dependence is removed by invoking Eq.\ \ref{eq:path}. In this step, we project $\phi$ onto $\phi(\varphi)$ and normalize to obtain
\begin{equation*}
\langle \varphi | \psi\rangle = \frac{\langle \varphi, \phi(\varphi) | \psi\rangle}{\sqrt{\int_0^{2\pi} \mathrm{d}\varphi \, |\langle \varphi, \phi(\varphi)| \psi\rangle|^2}}.
\end{equation*}
Finally, the Fourier transform of this function yields the charge wavefunctions
\begin{equation*}
\langle N |\psi\rangle = \frac{1}{2\pi} \int_0^{2\pi} \mathrm{d}\varphi \, \text{e}^{iN\varphi} \langle \varphi | \psi\rangle,
\end{equation*}
which are plotted in Fig.\ \ref{fig:spectrumb}.

\section{Parity sectors\label{app:sector}}

To explain the energy level structure at $\varphi_\text{ext} = \pi$, we factor Eq.\ \ref{eq:2m_ham} into the form \cite{Aasen2016}
\begin{equation*}
H_\text{eff} = H_+ \oplus H_-.
\end{equation*}
The parity sectors are governed by the Hamiltonians
\begin{align*}
H_\pm &= 4 \epsilon_C \left[ \frac{1}{4 (1-z)}(2\tilde{N} + k_\pm - N_\text{g} - \eta)^2 + x\eta^2\right] \nonumber \\
&\hspace{1cm} + \epsilon_L \theta^2 - \epsilon_J \left(1 - \frac{5}{4} z\right) \cos \tilde{\varphi}.
\end{align*}
In the above, $k_+ = 0$ and $k_- = 1$ while $\tilde{\varphi}$ and $\tilde{N}$ should be viewed as conjugate operators corresponding to pairs of Cooper pairs. Note that this expression is an exact reformulation of Eq.\ \ref{eq:2m_ham}. In the limit that $\epsilon_J \gg \epsilon_C$, we may Taylor expand the potential about $\tilde{\varphi} = 0$ and retain the first few terms. This step discards the effects of the offset charge, rendering $H_+$ and $H_-$ identical. We then decompose the linear part of the Hamiltonian into normal modes, which yields
%
%
%
\begin{align}
H_\pm &\approx 4 \epsilon_C \left( \frac{1}{1-z} \tilde{N}_n^2 + \frac{x}{1+z} \eta_n^2 \right) + \epsilon_L \theta_n^2 \nonumber \\
&\hspace{0.5cm}+\frac{1}{2}\epsilon_J \left(1 - \frac{3}{4}z\right)\tilde{\varphi}_n^2 - \frac{1}{24} \epsilon_J \left(\tilde{\varphi}_n + z\theta_n\right)^4 \label{eq:2m_ham_n}
\end{align}
to leading order in $z$. The corresponding transformation is given by
\begin{equation*}
\begin{pmatrix} \tilde{\varphi} \\[5pt] \theta \end{pmatrix} =
\begin{pmatrix} 1 & z \\[5pt] -\frac{1}{2} & 1 \end{pmatrix}
\begin{pmatrix} \tilde{\varphi}_n \\[5pt] \theta_n \end{pmatrix}.
\end{equation*}
Eq.\ \ref{eq:2m_ham_n} reveals two weakly anharmonic modes: the plasmon mode at the low frequency of $\sqrt{16 x \epsilon_L \epsilon_C}/h$ and a junction self-resonant mode at the high frequency of $\sqrt{8 \epsilon_J \epsilon_C}/h$. These modes are coupled by a quartic nonlinearity, which has the primary effect of inducing a Kerr shift on the junction self-resonant mode. At frequencies lower than $\sqrt{8 \epsilon_J \epsilon_C}/h$, the energy level structure is that of a two-fold degenerate harmonic oscillator, in agreement with the simulated energy spectrum at $\varphi_\text{ext} = \pi$ (see the dashed line in Fig.\ \ref{fig:spectruma}).

\section{Coherence estimates\label{app:estimates}}

\begin{figure*}
\centering
\subfloat{%
\includegraphics{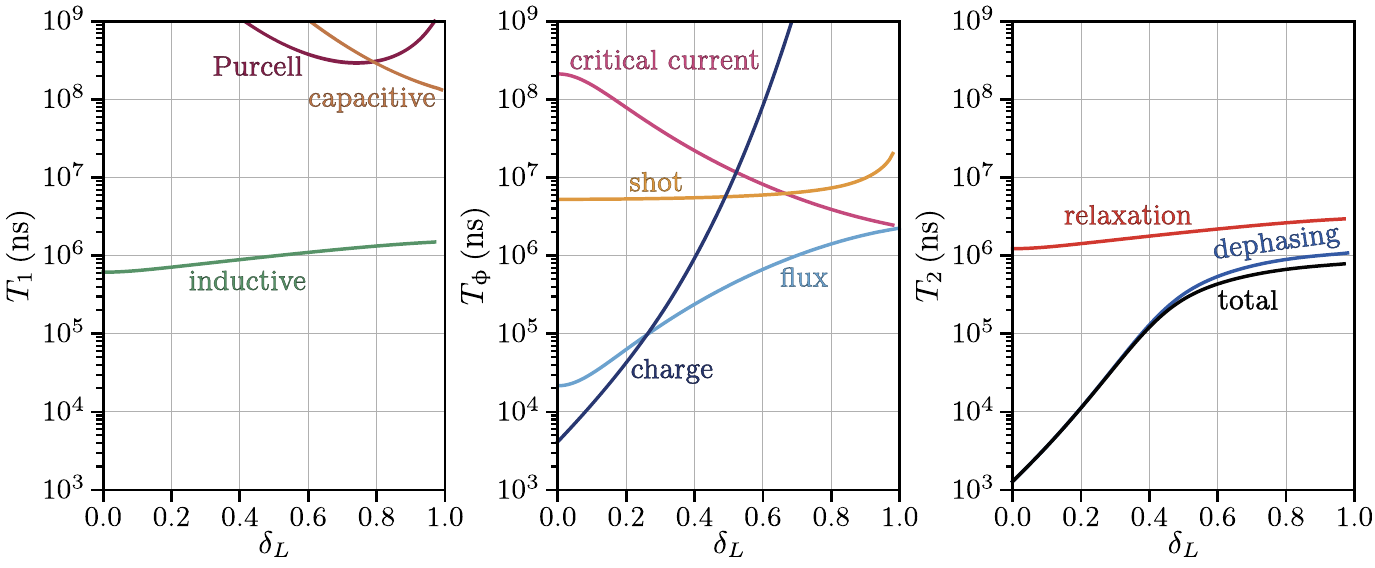}
\put(-398,157){(a)}
\put(-267,157){(b)}
\put(-135,157){(c)}
\label{fig:coherencea}}
\subfloat{
\label{fig:coherenceb}}
\subfloat{
\label{fig:coherencec}}
\caption{(a) Calculated relaxation times $T_1$ for the protected qubit as a function of inductive asymmetry $\delta_L$ (see Tab.\ \ref{tab:coherence}) for four different loss mechanisms. The expected contribution due to quasiparticle loss across the small junctions is too small to be seen on this plot. (b) Calculated pure dephasing times $T_\upphi$ for the qubit for four noise channels. (c) Combined estimates of the coherence time $T_2$ inferred using $1/T_2 = 1/2T_1 + 1/T_\upphi$. \label{fig:coherence}}
\end{figure*}

To expand on Secs.\ \ref{subsec:T1} and \ref{subsec:T2}, where relaxation pure dephasing times were calculated for four values of $\delta_L$, we plot in Fig.\ \ref{fig:coherence} the results for the entire span of $\delta_L$. In addition, in Fig.\ \ref{fig:coherencec} we also plot the expected relaxation, pure dephasing, and decoherence times combining the effects of all considered channels. Here, the overall decoherence time is defined using the relation $1/T_2 = 1/2T_1 + 1/T_\upphi$. We also note that, for the estimate of the dephasing rate due to charge noise, we do not use the extrapolation shown in Fig.\ \ref{fig:disordera} for the cases of disorder in $\epsilon_J$, $\epsilon_C$, and $\epsilon_A$. Numerical instabilities are not encounted for disorder in $\epsilon_L$ until $\delta_L \approx 0.95$ (which is outside the plotted range for both Fig.\ \ref{fig:disordera} and Fig.\ \ref{fig:coherenceb}).

In the remainder of this appendix, we describe the assumed frequency dependence of the three quality factors used for the relaxation time estimates in Sec.\ \ref{subsec:T1}. First, for the dielectric quality factor, we use the form \cite{Braginsky1987, Pop2014}
\begin{equation*}
\mathcal{Q}_\text{cap}(\omega) = Q_\text{cap} \bigg( \frac{2\pi \times 6 \, \text{GHz}}{|\omega|}\bigg)^{0.7}
\end{equation*}
so that the nominal value $Q_\text{cap} \sim 1\times 10^6$ corresponds to measurements performed at the resonant frequency of $6\,\text{GHz}$ \cite{Wang2015}.

Second, for quasiparticle loss, the dissipative part of the admittance for a Josephson junction with tunneling energy $\epsilon_J$ is given \cite{Catelani2011} to be
\begin{align}
\text{Re}\,Y_\text{qp}(\omega) &= \sqrt{\frac{2}{\pi}} \frac{8\epsilon_J}{R_\text{K} \Delta} \bigg(\frac{2\Delta}{\hbar\omega}\bigg)^{3/2} x_\text{qp} \sqrt{\frac{\hbar\omega}{2k_\text{B} T}} \nonumber \\
&\hspace{0.5cm} \times K_0 \bigg(\frac{\hbar|\omega|}{2k_\text{B} T}\bigg) \sinh \frac{\hbar\omega}{2k_\text{B} T}, \label{eq:yqp}
\end{align}
where $R_\text{K} = h/e^2$ is the resistance quantum, $\Delta$ is the superconducting gap, and $K_0$ is the modified Bessel function of the second kind. This expression holds under the assumption that the quasiparticle bath is in thermal equilibrium at temperatures $T \ll \Delta/k_\text{B}$, which may not be correct, depending on the implementation \cite{Serniak2018}. As we have seen, the matrix elements for quasiparticle loss vanish for this circuit, so this admittance has little bearing on our relaxation time estimates. 

On the other hand, inductive loss has been suggested to occur due to quasiparticle tunneling across Josephson junctions that compose superinductances \cite{Pop2014}. If we expect the frequency-dependence of the dissipative part of the admittance to agree between quasiparticle and inductive loss, then we arrive at
\begin{equation*}
\mathcal{Q}_\text{ind}(\omega) = Q_\text{ind} \frac{K_0 \big(\frac{h \times 0.5 \, \text{GHz}}{2 k_\text{B} T} \big) \sinh \big(\frac{h \times 0.5 \, \text{GHz}}{2 k_\text{B} T} \big)}{K_0 \big(\frac{\hbar|\omega|}{2k_\text{B} T}\big) \sinh \big(\frac{\hbar|\omega|}{2k_\text{B} T}\big)}
\end{equation*}
so that the nominal value $Q_\text{ind} \sim 500 \times 10^6$ corresponds to measurements performed at the resonant frequency of $0.5\,\text{GHz}$.

\bibliography{thesis_bibtex}

\end{document}